\documentclass[12pt]{article}
\usepackage{amsfonts,amssymb}
\usepackage{color}
\usepackage{epic}
\usepackage{graphics}

\def\lddots{\mathinner{\mkern1mu\raise1pt\hbox{.}\mkern2mu
\raise4pt\hbox{.}\mkern2mu\raise7pt\vbox{\kern7pt\hbox{.}}\mkern1mu}}
\makeatletter
\def\numberbysection{\@addtoreset{equation}{section}
\def\theequation{\thesection.\arabic{equation}}}
\makeatother

\numberbysection

\newcommand{\be}{\begin{eqnarray}}
\newcommand{\ee}{\end{eqnarray}}
\newcommand{\non}{\nonumber}

\newcommand{\tr}{\mathop{\rm tr}\nolimits}

\begin{document}

\begin{titlepage}
\vskip 0.4cm
\strut\hfill
\vskip 0.8cm
\begin{center}

%\begin{center}

{\bf {\Large Boundary Lax pairs from non-ultra local Poisson algebras}}

\vspace{10mm}

{\bf Jean Avan\footnote{avan@ptm.u-cergy.fr}$^{a}$}\phantom{x} and\phantom{x}
{\bf Anastasia Doikou\footnote{adoikou@upatras.gr}$^{b}$}

\vspace{10mm}

{\small $^a$ LPTM, Universite de Cergy-Pontoise (CNRS UMR 8089), Saint-Martin 2\\
2 avenue Adolphe Chauvin, F-95302 Cergy-Pontoise Cedex, France}

{\small $^b$ University of Patras, Department of Engineering Sciences,\\
GR-26500, Patras, Greece}

\end{center}

\vfill

\begin{abstract}

We consider non-ultra local linear Poisson algebras on a continuous line. Suitable combinations
of representations of these algebras yield representations of novel generalized linear Poisson
algebras or ``boundary'' extensions. They are parametrized by a ``boundary'' scalar matrix and
depend in addition on the choice of an anti-automorphism. The new algebras are the classical-linear
counterparts of known quadratic
quantum boundary algebras. For any choice of parameters the non-ultra local contribution of
the original Poisson algebra disappears. We also systematically construct the associated classical
Lax pair. The classical boundary PCM model is examined as a physical example.

\end{abstract}

\vfill
\baselineskip=16pt

\end{titlepage}

\section{Introduction}

We have recently derived a general algebraic construction for the Lax representation
of a classical integrable system when the Lax matrix obeyed a
Poisson bracket algebra identified with several so-called boundary algebras \cite{avandoikou}.
The systematic formulation of the boundary algebra and Lax pair, given an original
classical $r$-matrix structure, gives to our derivation a universal
nature which we expect makes it very useful to define and study classical
integrable systems with boundaries (see examples in \cite{avandoikou1})

We had previously examined the situation where the Poisson structure for the classical
Lax operator was expressed in terms of a single non-dynamical parametric
skew-symmetric classical $r$-matrix together with a set of $c$-number
non-dynamical parameters encapsulated in a single $k$ matrix obeying
a purely algebraic quadratic equation expressed in term of the $r$ matrix.
This formulation was in fact a classical limit of the quantum reflection
algebra of Cherednik and Sklyanin \cite{sklyanin} symbolically written
$R_{12}K_1R'_{12}K_2 = K_2R'_{12}K_1R_{12}$ where e.g. $R'_{12} = R_{21}$

The most general reflection equation, derived by Freidel and Maillet \cite{maillet},
is a $4$-matrix quadratic structure $A_{12}K_1B'_{12}K_2 = K_2C'_{12}K_1D_{12}$,
together with some relation between $A,B,C,D$ guaranteeing the existence of an invertible
scalar $k$ matrix solution, the simplest one being $AB = CD$; additional
unitarity conditions require for instance  $B_{12} = C_{21}$. It has a classical limit
involving a non-skew symmetric $r$ matrix $r = a+b$ (lower case letters denote
respective classical limits of quantum matrices $ABCD$. Such classical
limits allow \cite{maillet} for construction in particular of non-ultralocal
linear Poisson algebra structures.

We shall present here the generalization of \cite{avandoikou} to such a situation
where in addition the partner matrix $b$
in the ``pair'' of $r$ matrices is a symmetric matrix denoted by $s$. (The most
general situation where $b$ has no particular property will be left for
further studies).

We shall first establish the general Poisson bracket structure for the local relevant combinations of Lax matrix and boundary
matrix (with suitable constraints) relevant in this context. It will be noted that
the $s$ contribution vanishes from the boundary algebra, at least such as is defined here.

We shall then consider the example of an integrated boundary algebra constructed from
the $r,s$ pair of the classical Principal Chiral Model \cite{maillet}.

By examining this particular example we show that the boundary non-local charges emerging from this model
are in exact correspondence with the quantum boundary non-local charges derived in earlier works
(see e.g \cite{doikouy}), for various types of boundary conditions. In addition these models seem to play an important role within
the AdS/CFT context (see e.g. \cite{ads}) especially from the string theory point of view, whereas their discrete quantum counterparts
should be of great significance from the gauge theory point of view.

An integrated version starting from the local Poisson algebra is then considered, relying
on the construction of a suitable monodromy matrix, obeying now quadratic Poisson algebras. The generating matrix
is obtained by computing a monodromy matrix for the non ultra-local Lax matrix, then twisting
it by an appropriate automorphism. The reverse order of procedure will not be considered
here. Again the $s$ contribution
vanishes after integration, in a way recalling the computation of the equal-interval
Poisson structure for the monodromy matrix in the complex sine Gordon case \cite{maillet3}.
The non-local classical charges are then constructed following
our general scheme; they are in exact agreement with the previously computed quantum
non local charges \cite{doikouy}, thereby validating the whole approach for our classical
reflection algebras in this generalized, non skew-symmetric non ultra-local framework.

\section{Linear reflection Poisson algebras from $r,s$ pairs}

Let us consider the $L$ matrix \cite{lax} satisfying the fundamental
non-ultra linear local Poisson algebra \cite{maillet}: \be \Big
\{L(\lambda, x),\otimes \ \ L(\mu, y) \Big \} &=& \delta(x-y) \Big (
\Big [ (r-s)(\lambda, \mu),\ L(\lambda, x)\otimes {\mathbb I}   \Big
] \non\\ &+&  \Big [ (r+s)(\lambda, \mu),\ {\mathbb I}  \otimes
L(\mu, x)   \Big ] \Big ) - 2 s(\lambda, \mu)\delta'(x-y)
\label{fundam} \ee where the associativity condition for the Poisson
bracket yield sufficient conditions for the pair $r,s$ expressed as
the general classical Yang-Baxter equation \cite{skl, sts}: \be &&
[(r+s)_{13}(\lambda, \eta),\ (r-s)_{12}(\lambda, \mu)] +
[(r+s)_{23}(\mu, \eta),\ (r+s)_{12}(\lambda, \mu)] \non\\ &&+
[(r+s)_{23}(\lambda, \mu),\ (r+s)_{13}(\lambda, \eta)] =0. \ee Here
$r$ is skew-symmetric: $r_{12} = -r_{21}$; $s$ is symmetric: $s_{12}
= s_{21}$. This choice of decomposition of a non-skew-symmetric
$r$-matrix $r+s$ is quite general although not universal: there
exists for instance situations where the classical $r$ matrix is
more relevantly decomposed into a skew-symmetric part and a
semi-diagonal part (see e.g. the $r$ matrix structure of the scalar
Calogero-Moser Lax matrix) but we shall not consider this situation
here.

Inspired essentially by the already constructed boundary type algebras \cite{avandoikou}
we define the following object. Let $\sigma$ be any anti-automorphism of the Lie
algebra (\ref{fundam}), then define:
\be
{\mathbb T}(\lambda, x) = L(\lambda,x)\ {\mathrm k}(\lambda) + {\mathrm k}(\lambda)\ L^{\sigma}(\lambda,x). \label{rep}
\ee
We show that the Poisson brackets of ${\mathbb T}$ are expressed in
certain algebraic forms subject to sets of consistency conditions on ${\mathrm k}$.
There are in particular two specific anti-automorphisms with an interesting
physical interpretation, which have been extensively
discussed in the context of integrable boundary conditions: they describe the so-called
soliton-preserving (SP) boundary conditions associated to the reflection algebra and
the soliton-non preserving (SNP) associated to the twisted Yangian (for a
detailed discussion and relevant references see e.g. \cite{avandoikou, avandoikou1}).
For historical reasons we shall keep the terminology `reflection algebras' and `twisted
Yangians' for the linear algebras associated to these two physical anti-automorphisms defined below:
\be
&& L^{\sigma}(\lambda) = -L(-\lambda) ~~~~\mbox{Reflection algebra},
\non\\ && L^{\sigma}(\lambda) = L^t(-\lambda) ~~~~\mbox{Twisted Yangian}.
\label{anti}
\ee

We now derive  the Poisson bracket: $\{ {\mathbb T}(\lambda), \otimes \ {\mathbb T}(\mu)\}$.
To achieve this we shall need in addition to (\ref{fundam}) the following exchange relations,
which follow from (\ref{fundam}) under the action of the anti-morphism:
\be
\Big
\{ L^{\sigma} (\lambda, x), \otimes\  L(\mu, y)  \Big \} &=& \delta(x-y)\Big (-\Big
[(r-s)^{\sigma_1}(\lambda,\mu),\ L^{\sigma}(\lambda, x)\otimes {\mathbb I} \Big ] \non\\
&+& \Big [(r+s)^{\sigma_1}(\lambda,\mu),\ {\mathbb I}\otimes L(\mu,x)\Big ] \Big )  -
2\delta'(x-y)s^{\sigma_1}(\lambda, \mu)
\non\\  \Big \{ L(\lambda, x), \otimes\  L^{\sigma}(\mu, y)  \Big \} &=& \delta(x-y)\Big
(\Big [(r-s)^{\sigma_2}(\lambda,\mu),\ L(\lambda, x) \otimes {\mathbb I}\Big ] \non\\ &-&
\Big [(r+s)^{\sigma_2}(\lambda,\mu),\ {\mathbb I}\otimes L^{\sigma}(\mu, x)\Big ] \Big )  -
2 \delta'(x-y) s^{\sigma_2}(\lambda, \mu)
\non\\
\Big \{ L^{\sigma}(\lambda, x), \otimes\  L^{\sigma}(\mu, y)  \Big \} &=& \delta(x-y)\Big
(-\Big [(r-s)^{\sigma_1 \sigma_2}(\lambda,\mu),\ L^{\sigma}(\lambda)\otimes {\mathbb I}\Big ]
\non\\ &-& \Big [(r+s)^{\sigma_1 \sigma_2}(\lambda,\mu),\ {\mathbb I} \otimes L^{\sigma}(\mu,x)\Big ] \Big )
-2\delta'(x-y)s^{\sigma_1 \sigma_2}(\lambda, \mu). \label {fundam2}
\ee

By explicit use of the algebraic relations (\ref{fundam}) and (\ref{fundam2}) we obtain (recall the notation
$A_1 = A \otimes {\mathbb I},\ A_2 = {\mathbb I} \otimes A$):
\be
&& \Big \{ {\mathbb T}_1(\lambda, x),  \ {\mathbb T}_2(\mu, y)\ \Big \}= \Big \{L_1(\lambda, x){\mathrm k}_1(\lambda)
+ \hat L_1(\lambda, x) {\mathrm k}_1(\lambda) ,\ L_2(\mu,y) {\mathrm
k}_2(\mu)+ {\mathrm k}_2(\mu)\hat L_2(\mu, y) \Big \} = \ldots \non\\
& =& \delta(x-y) \Big ( \Big ((r-s)_{12}(\lambda, \mu){\mathrm k}_2(\mu) +  {\mathrm k}_2(\mu) (r-s)_{12}^{\sigma_2}(\lambda, \mu)) \Big )
L_1(\lambda, x) {\mathrm k}_1(\lambda)  \non\\ &+&  \Big (-{\mathrm k}_1(\lambda) {\mathrm k}_2(\mu) (r-s)_{12}^{\sigma_1 \sigma_2}
(\lambda, \mu)-
{\mathrm k}_1(\lambda) (r-s)_{12}^{\sigma_1}(\lambda, \mu){\mathrm k}_2(\mu) {\mathrm k}_1^{-1}(\lambda) \Big ) {\mathrm k}_1(\lambda)
L^{\sigma}_1(\lambda,x) \non\\
&-& L_1(\lambda,x) {\mathrm k}_1(\lambda) \Big ({\mathrm k}_1^{-1}(\lambda) (r-s)_{12}^{\sigma_1 \sigma_2}(\lambda, \mu ){\mathrm k_1}
(\lambda)
{\mathrm k_2}(\mu) +{\mathrm k}_1^{-1}(\lambda) {\mathrm k}_2(\mu)(r-s)_{12}^{t_2}(\lambda, \mu) \Big ) \non\\ &-& {\mathrm k}_1(\lambda)
L^{\sigma}_1 (\lambda, x)\Big (-{\mathrm k}_2(\mu) (r-s)_{12}^{\sigma_1 \sigma_2}(\lambda, \mu) - (r-s)_{12}^{\sigma_1}(\lambda, \mu)
{\mathrm k}_2(\mu)
\Big ) \non\\ &+& \Big ((r+s)_{12}(\lambda, \mu){\mathrm k}_1(\lambda) + {\mathrm k}_1(\lambda)(r+s)^{\sigma_1}(\lambda, \mu) \Big )
L_2(\mu,y)
{\mathrm k}_2(\mu) \non\\ &+& \Big (-{\mathrm k}_1(\lambda) {\mathrm k}_2(\mu)(r+s)_{12}^{\sigma_1 \sigma_2}(\lambda, \mu) -
{\mathrm k}_2(\mu)
(r+s)_{12}^{\sigma_2}
(\lambda, \mu){\mathrm k}_1(\lambda){\mathrm k}_2^{-1}(\mu) \Big ) {\mathrm k}_2(\mu) L^{\sigma}(\mu, y)\non\\ &-& L_2(\mu, y)
{\mathrm k}_2(\mu)
\Big ({\mathrm k}_2^{-1}(\lambda) (r+s)_{12}(\lambda, \mu){\mathrm k}_1(\lambda){\mathrm k}_2(\mu) + {\mathrm k}_2^{-1}(\mu){\mathrm k}_1
(\lambda)(r+s)_{12}^{\sigma_1}
(\lambda, \mu) {\mathrm k}_2(\mu) \Big)\non\\ &-& {\mathrm k}_2(\mu) L^{\sigma}_2(\mu, y)\Big (-{\mathrm k}_1(\lambda)(r+s)_{12}
^{\sigma_1 \sigma_2}(\lambda, \mu) - (r+s)_{12}^{t_2}(\lambda, \mu){\mathrm k}_1(\lambda) \Big )\Big ) \non\\ &-& 2 \delta'(x-y) \Big
( s_{12}(\lambda,\mu) {\mathrm k}_1(\lambda)
{\mathrm k}_2(\mu)+ {\mathrm k}_1(\lambda) {\mathrm k}_2(\mu) s_{12}^{\sigma_1 \sigma_2}(\lambda,\mu) \non\\ && +\ {\mathrm k}_1(\lambda)
s_{12}^{\sigma_1}(\lambda, \mu){\mathrm k}_2(\mu) +
{\mathrm k}_2(\mu) s_{12}^{\sigma_2}(\lambda, \mu) {\mathrm k}_1(\lambda)\Big ). \non\\ \label{fin}
\ee

By imposing the following set of constraints on ${\mathrm k}$
\be
&& {\mathrm A}_{12}(\lambda,\mu)\ {\mathrm k}_1(\lambda)\ {\mathrm k}_2(\mu)+ {\mathrm k}_1(\lambda)\ {\mathrm k}_2(\mu)\
{\mathrm A}_{12}^{\sigma_1 \sigma_2}(\lambda,\mu) \non\\ && +\
{\mathrm k}_1(\lambda)\ {\mathrm A}_{12}^{\sigma_1}(\lambda, \mu)\ {\mathrm k}_2(\mu) + {\mathrm k}_2(\mu)\
{\mathrm A}_{12}^{\sigma_2}(\lambda, \mu)\ {\mathrm k}_1(\lambda) =0,
\non\\
&& \mbox{where} ~~~~~{\mathrm A}_{12}(\lambda, \mu) \equiv r(\lambda, \mu), \ s(\lambda, \mu)
\label{re1}
\ee
we end up with the generalized linear algebraic relations:
\be
&& \Big \{ {\mathbb T}_1(\lambda,x),\ {\mathbb T}_2(\mu, y) \Big \} = \non\\ && \Big (
{\mathrm r}_{12}^{-} (\lambda, \mu) {\mathbb T}_1(\lambda, x) - {\mathbb
T}_1(\lambda, x) \tilde {\mathrm r}_{12}^{-} (\lambda, \mu) +
{\mathrm r}_{12}^{+} (\lambda, \mu){\mathbb T}_2(\mu, y)-{\mathbb T}_2(\mu, y)
\tilde {\mathrm r}_{12}^{+} (\lambda, \mu) \Big )\delta(x-y) \non\\  \label{gen}
\ee
where we define\footnote{Note that constraints (\ref{re1}) may
be equivalently expressed as
\be
&& {\mathrm k}_1^{-1}(\lambda)
{\mathrm r}^-_{12}(\lambda, \mu) -\tilde {\mathrm r}^-_{12}(\lambda, \mu)
{\mathrm k}^{-1}_1(\lambda) =0 \non\\ && {\mathrm k}_2^{-1}(\mu) {\mathrm r}^+_{12}(\lambda, \mu)
-\tilde {\mathrm r}^+_{12}(\lambda, \mu) {\mathrm k}^{-1}_2(\mu) =0 \ee} \be {\mathrm r}_{12}^
-(\lambda, \mu) &=& (r-s)_{12}(\lambda, \mu){\mathrm k}_2(\mu) +{\mathrm k}_2(\mu) (r-s)_{12}^{\sigma_2}
(\lambda, \mu), \non\\ \tilde {\mathrm r}_{12}^-(\lambda, \mu) &=& -{\mathrm k}_2(\mu)(r-s)_{12}^{\sigma_1 \sigma_2}(\lambda, \mu)
- (r-s)_{12}^{\sigma_1}(\lambda, \mu){\mathrm k}_2(\mu) \non\\ {\mathrm r}^+_{12}(\lambda, \mu)&=& (r+s)_{12}(\lambda, \mu){\mathrm k}_1
(\lambda)
+ {\mathrm k}_1(\lambda) (r+s)_{12}^{\sigma_1}(\lambda, \mu) \non\\  \tilde {\mathrm r}^+_{12}(\lambda, \mu) &=& - {\mathrm k}_1(\lambda)
(r+s)_{12}^{\sigma_1 \sigma_2}(\lambda, \mu)- (r+s)_{12}^{\sigma_2}(\lambda, \mu){\mathrm k}_1(\lambda). \label{re2}
\ee

A significant consequence of the algebraic construction above
is that the non-ultra locality of the original linear algebra (\ref{fundam})
is lost no matter what matrix ${\mathrm k}$ we
choose, as long as it satisfies the constraint (\ref{re1})!

Define now  ${\mathbb L}(\lambda,x) = {\mathrm k}^{-1}(\lambda)\ {\mathbb
T}(\lambda,x)$, we prove:
\be
\Big \{ tr_a {\mathbb
L}_a^N(\lambda,x),\ tr_b {\mathbb L}^M_b(\mu,y) \Big \} = 0. \label{th2} \ee
Indeed from: \be && \Big \{ tr_a {\mathbb
L}_a^N(\lambda,x),\ tr_b {\mathbb L}^M_b(\mu,y) \Big \} =\non\\ && \sum_{n, m}
tr_{ab}\ {\mathbb L}_a^{N-n}(\lambda,x) {\mathbb L}_b^{M-m}(\mu,y)
\Big \{ {\mathbb L}_a(\lambda,x),\ {\mathbb L}_b(\mu,y) \Big \}
{\mathbb L}_a^{n-1}(\lambda,x) {\mathbb L}_b^{m-1}(\mu,y) \ee
employing (\ref{gen}) the preceding expression becomes \be && \dots
\propto \non\\ && tr_{ab} \Big \{ {\mathbb L}^{N-1}_a(\lambda,x) {\mathbb
L}^{M-1}_{b}(\mu,y) {\mathrm k}_a^{-1}(\lambda) {\mathrm k}_b^{-1}(\mu) \non\\ && \Big (
{\mathrm r}_{ab}^{-} (\lambda, \mu) {\mathbb T}_a(\lambda,x) - {\mathbb
T}_a(\lambda,x) \tilde {\mathrm r}_{ab}^{-} (\lambda, \mu) +
{\mathrm r}_{ab}^{+} (\lambda, \mu){\mathbb T}_b(\mu,y)-{\mathbb T}_b(\mu,y)
\tilde {\mathrm r}_{ab}^{+} (\lambda, \mu) \Big ) \Big \} \delta(x-y) \non\\ &&=
\ldots =0.
\ee
Note that in order to show that the latter
expression is zero we moved appropriately the factors in the
products within the trace and we used (\ref{gen}).

We finally identify the Lax formulation associated
(see also \cite{avandoikou, BBT, babelon})
to the generalized algebra (\ref{re1})--(\ref{re2}).
Defining Hamiltonians as: $tr_a {\mathbb L}^n(\lambda) =
\sum_i{{\cal H}_n^{(i)} \over \lambda^i}$ the
classical equations of motion for ${\mathbb L}$:
\be
{\dot {\mathbb
L}}(\mu,y)= \Big \{{\cal H}_n^{(i)},\ {\mathbb L}(\mu,y) \Big \}
\label{cleq}
\ee
take a zero
curvature form
\be
{\dot {\mathbb L}}(\mu,y) =  \Big [{\mathbb A}(\lambda, \mu,x),\
{\mathbb L}(\mu,y) \Big], \ee where ${\mathbb A}_n$ is
identified as: \be {\mathbb A}_n(\lambda, \mu,x)  =n\  tr_a \Big (
{\mathbb L}_a^{n-1}(\lambda,x) {\mathrm k}_a^{-1}(\lambda) \tilde
{\mathrm r}^+_{ab}(\lambda, \mu) \Big ). \label{final3}
\ee

The proof runs as follows.

Consider
\be
&& \Big
\{tr_a {\mathbb L}^n_a(\lambda,x),\ {\mathbb L}_b(\mu,y) \Big \}
=\ldots= \non\\ && n\ tr_a \Big ( {\mathbb L}_a^{n-1}(\lambda,x)
{\mathrm k}_a^{-1}(\lambda) {\mathrm k}_b^{-1}(\mu) ({\mathrm
r}^-_{ab}(\lambda, \mu) {\mathbb T}_a(\lambda,x) -{\mathbb
T}_a(\lambda,x)\tilde {\mathrm r}^-_{ab}(\lambda,\mu) \non\\ && +{\mathrm
r}^+_{ab}(\lambda, \mu) {\mathbb T}_b(\mu,y) -{\mathbb T}_b(\mu,y) \tilde
{\mathrm r}^+_{ab}(\lambda,\mu)) \Big )\delta(x-y) \non\\ &&= n\ tr_a \Big (
{\mathbb L}_a^{n-1}(\lambda,x) {\mathrm k}_a^{-1}(\lambda) {\mathrm
k}_b(\mu)({\mathrm r}^{-}_{ab}(\lambda, \mu){\mathbb T}_a(\lambda,x)
-{\mathbb T}_a(\lambda,x) \tilde {\mathrm r}^{-}_{ab}(\lambda, \mu) )
\Big ) \delta(x-y) \non\\ && + n\ tr_a\Big ({\mathbb L}_a^{n-1}(\lambda,x) {\mathrm
k}_a^{-1}(\lambda) {\mathrm k}_b(\mu)({\mathrm
r}^{+}_{ab}(\lambda, \mu){\mathbb T}_b(\mu,y) -{\mathbb T}_b(\mu,y) \tilde
{\mathrm r}^{+}_{ab}(\lambda, \mu))  \Big ) \delta(x-y). \non\\
\ee
Taking into account (\ref{gen}) we see that the first term of RHS of the
equality above disappears and the last term may be appropriately
rewritten as:
\be
\Big \{tr_a {\mathbb L}^n_a(\lambda,x),\
{\mathbb L}_b(\mu,y) \Big \} &=& \Big (n\ tr_a \Big ({\mathbb
L}^{n-1}(\lambda,x){\mathrm k}_a^{-1}(\lambda) \tilde {\mathrm
r}^{+}_{ab}(\lambda, \mu) \Big )\ {\mathbb L}_{b}(\mu,y) \non\\ &-& n\
{\mathbb L}_{b}(\mu,x)\  tr_a \Big ( {\mathbb L}^{n-1}(\lambda,x)
{\mathrm k}_a^{-1}(\lambda) \tilde {\mathrm r}^{+}_{ab}(\lambda,
\mu) \Big )\Big )\delta(x-y). \non\\
\ee
From the latter formula (\ref{final3}) is finally
deduced.

\section{Application: the Principal Chiral model}

Rather than considering the general setting for derivation of the integrated version
of our algebraic structure by construction of a suitable monodromy matrix we shall
exemplify it on the physically relevant case of the Principal Chiral Model algebra.

\subsection{The local Poisson structure}

The $r,\ s$ matrices are chosen to be (see e.g. \cite{maillet, young}):
\be
r(\lambda, \mu) &=& {1\over 2(\lambda - \mu)} \Big ( {\mu^2 \over \mu^2 -1} + {\lambda^2 \over \lambda^2 -1} \Big )\
\Pi \non\\ s(\lambda, \mu) &=& - {(\lambda + \mu) \over 2 (\lambda^2-1)(\mu^2-1)}\ \Pi \label{rr}
\ee
and are solutions of the classical Yang-Baxter type equation.

The particular Lax operator (see e.g. \cite{ft, maillet} and references therein) here takes the form:
\be
L(\lambda,x ) = {\lambda j_{0}(x) + j_{1}(x) \over \lambda^2 -1}
\ee
It satisfies (\ref{fundam}). Consider the generators of the algebra ${\mathfrak g}$ satisfying
\be
[t_a,\ t_b] =
C_{abc}\ t_c
\ee
where $C_{abc}$ is a fully antisymmetric tensor. Then the local
currents $j_\mu(x)$, $\mu \in \{0,\ 1 \}$ may be expressed as:
\be
j(x) = \sum_a j^a(x) t_a.
\ee
Define also
\be
\Pi = \sum_a t_a \otimes t_a
\ee
which satisfies
\be
[\Pi,\ t_c\otimes {\mathbb I}]= - [\Pi,\ {\mathbb I} \otimes t_c ] = C_{bca}\ t_a\otimes t_b.
\ee
In the defining representation of $gl_n$ for instance it is easy to show that $\Pi \propto {\cal P} + c$, recalling that
${\cal P}$ is the permutation operator and  $c$-depends on the rank of the algebra.

The currents $j_i$ indeed obey:
\be
\Big \{j_{0}(x),\otimes \ j_{0}(y) \Big \} &=& \delta(x-y) \Big [\Pi,\
{\mathbb I} \otimes j_{0}(x)   \Big ] \non\\   \Big \{j_{0}(x), \otimes \ j_{1}(y) \Big \} &=& \delta(x-y)\Big [\Pi,\
{\mathbb I} \otimes j_{1}(x) \Big ] + 2\ \Pi\ \delta'(x-y)\non\\
\Big \{j_{1}(x),\otimes \ j_{1}(y) \Big \} &=& 0.
\ee
We shall focus henceforth on the two physical anti-morphisms (\ref{anti}).
Then the `boundary' currents are defined from ${\mathbb T}$ (\ref{rep}) in a straightforward manner as --note that here we consider
for simplicity constant ($\lambda$-independent) ${\mathrm k}$ matrices:
\be
j_{0}^{(b)}(x) =j_{0}(x)\ {\mathrm k} +{\mathrm k}\ j_{0}(x), ~~~~~j_{1}^{(b)}(x) =j_{1}(x)\ {\mathrm k} -{\mathrm k}\ j_{1}(x)
~~~~~~\mbox{Reflection algebra} \non\\
\tilde j_{0}^{(b)}(x) =\tilde j_{0}(x)\ \tilde {\mathrm k} -\tilde {\mathrm k}\
\tilde j_{0}(x)^{t}, ~~~~~\tilde j_{1}^{(b)}(x) =\tilde j_{1}(x)\ \tilde {\mathrm k} +\tilde {\mathrm k}\ \tilde j_{1}^{t_1}(x)
~~~~~~\mbox{Twisted Yangian}
\ee
and they satisfy the following algebraic relations.
\\
\\
{\it Reflection algebra:}
\be
\{ j^{(b)}_{0}(x), \otimes \ j^{(b)}_{\alpha}(y)\}&=& [{\mathrm r},\ {\mathbb I} \otimes j^{(b)}_{\alpha}(x) ] \delta(x-y),
~~~~~\alpha \in \{0,\ 1 \}, \non\\ \{j^{(b)}_{1}(x), \otimes \ j^{(b)}_{1}(y)\} &=& 0. \label{11}
\ee
\be
{\mathrm r}_{12} =  \Pi_{12}  {\mathrm k}_1  +  {\mathrm k}_1 \Pi_{12}.
\ee
{\it Twisted Yangian:}
\be
\{\tilde j^{(b)}_{0}(x), \otimes \ \tilde j^{(b)}_{\alpha}(x)\}&=& [\tilde {\mathrm r},\ {\mathbb I} \otimes
\tilde j^{(b)}_{\alpha}(x) ] \delta(x-y),~~~~~\alpha \in \{0,\ 1 \}, \non\\ \{\tilde j^{(b)}_{1}(x), \otimes \
\tilde j^{(b)}_{1}(y)\} &=& 0 \label{22}
\ee
where we define
\be \tilde  {\mathrm r}_{12} =  \Pi_{12} \tilde {\mathrm k}_1   - \tilde {\mathrm k}_1 \Pi_{12}^{t_1}.
\ee
The exchange relations (\ref{11}), (\ref{22}) are valid provided that the ${\mathrm k}$ and $\tilde {\mathrm k}$
matrices satisfy the following constraints obtained from (\ref{re1})
\be  {\mathrm k}^2 \propto {\mathbb I}, ~~~~\tilde {\mathrm k}_1 \Pi_{12}^{t_1} \tilde {\mathrm k}_2 = \tilde {\mathrm k}_2
\Pi_{12}^{t_1}
\tilde {\mathrm k}_1. \label{restr} \ee
Let us consider the integrals of motion emerging from $tr{\mathbb L}^2$
\be {\mathbb L}(\lambda, x) =  {\mathrm k}^{-1} L(\lambda, x) {\mathrm k} -L(-\lambda, x) =
{1\over \lambda^2 -1} \Big (\lambda j_{0}^{(b)}(x)
+j_{1}^{(b)}(x) \Big ) \label{restr1}\ee
then
\be
\tr{\mathbb L}^2(\lambda,x) = \sum_k {{\mathbb I}^{(k)} \over \lambda^{k+1}}
\ee
and the explicit form
of the associated integrals of motion are:
\be
{\mathbb I}^{(1)}=\tr\Big ((j_{0}^{(b)}(x))^2\Big ), ~~~~~{\mathbb I}^{(2)}= 2 \tr\Big
(j_{0}^{(b)}(x)j_{1}^{(b)}(x)\Big ), ~~~~{\mathbb I}^{(3)} = \tr\Big ((j_{0}^{(b)}(x))^2 + (j_{1}^{(b)}(x))^2 \Big ),~~~ \ldots
\ee

The associated $A$ operators of the Lax pairs are then given by (\ref{final3}):
\be
{\mathbb A}_2(\lambda, \mu) = 2 \tr_a\{{\mathbb L}_{a}(\lambda, x) {\mathrm k}_a^{-1} \tilde {\mathrm r}_{ab}(\lambda, \mu)\}.
\ee

\subsection{The monodromy matrix and the boundary PCM model}

Having constructed the {\it local} boundary current algebra naturally
associated to the PCM model, we are now in position to construct
and study explicit models with open integrable boundary conditions by
deriving an ``integrated'' monodromy matrix, which will now be ruled by a quadratic boundary algebra.
We shall identify some of the boundary non-local charges, which belong to the quadratic algebra, by appropriately
expanding the associated generating function.

Note that in this approach we shall construct the final monodromy matrix in two
distinct steps: {\it first} integrate over the original Lax matrix $L$ obeying the
non-ultralocal Poisson structure (\ref{fundam}); {\it then} twisting the resulting
monodromy matrix. Reversing these operations should give a qualitatively different
algebraic structure and associated conserved quantities, and shall be discussed
in a later work.

We introduce the natural object:
\be
T(\lambda)\ = {\cal P} \exp\{\int_{-L}^{0} L(\lambda, x)\}. \label{mono2}
\ee
One can show in this particular case, taking into account the form of the $r,\ s$ matrices that $T$ satisfies
(see also \cite{maillet, ft, rajev}):
\be
\Big \{T_1(\lambda),\  T_2(\mu) \Big \} = \Big [r_{12}(\lambda,\mu),\ T_1(\lambda) T_2(\mu) \Big ],
\label{qua}
\ee
where $r$ is the matrix defined in (\ref{rr}). This result can be compared to the one in \cite{maillet3} where the Poisson
structure of the equal-point limit of the
monodromy matrix for the (continuous) complex sine Gordon model was shown to take
the form (\ref{qua}) even though the local Poisson brackets were of non-ultralocal form.

Henceforth, we shall consider two `physical' anti-automorphisms that correspond to two
distinct sets of boundary conditions (SP and SNP) associated with the reflection algebra and twisted Yangian respectively:
\be
&& T^{\sigma}(\lambda) = T^{-1}(-\lambda) ~~~\mbox{Reflection algebra},
\non\\ && T^{\sigma}(\lambda) = T^t(-\lambda) ~~~\mbox{Twisted Yangian}. \label{sigma2}
\ee
We define below the classical versions of the reflection algebra and the twisted Yangian (see e.g.
\cite{maillet, sklyanin}) associated to these two types of boundary conditions:
\\
\\
{\bf Reflection algebra (SP)}
\be
\Big \{{\cal T}_1(\lambda_1),\ {\cal
T}_2(\lambda_2) \Big \} &=& r_{12}(\lambda_1-\lambda_2){\cal
T}_{1}(\lambda_1){\cal T}_2(\lambda_2) -{\cal T}_1(\lambda_1)
{\cal T}_2(\lambda_2) r_{21}(\lambda_1 -\lambda_2) \non\\ & +&
{\cal T}_{1}(\lambda_1) r_{21}(\lambda_1+\lambda_2){\cal
T}_2(\lambda_2)- {\cal T}_{2}(\lambda_2)
r_{12}(\lambda_1+\lambda_2){\cal T}_1(\lambda_1). \label{refc1}
\ee
\\
{\bf Twisted Yangian (SNP)}
\be
\Big \{ \tilde {\cal T}_1(\lambda_1),\ \tilde {\cal
T}_2(\lambda_2) \Big \} &=& r_{12}(\lambda_1-\lambda_2)\tilde {\cal T}_{1}(\lambda_1)\tilde
{\cal T}_2(\lambda_2) -\tilde {\cal
T}_1(\lambda_1)
\tilde {\cal T}_2(\lambda_2) r^{t_1 t_2}_{21}(\lambda_1 -\lambda_2) \non\\
& +& \tilde {\mathbb T}_{1}(\lambda_1)
r_{12}^{t_1}(\lambda_1+\lambda_2)\tilde {\cal T}_2(\lambda_2)- \tilde {\cal
T}_{2}(\lambda_2) r_{21}^{t_2}(\lambda_1+\lambda_2)\tilde {\cal
T}_1(\lambda_1). \label{refc2}
\ee

Representations of the algebras above are given as is well known by:
\be
&& {\cal T}(\lambda) =
T(\lambda)\ K(\lambda)\ T^{-1}(-\lambda),
 \non\\  && \tilde {\cal T}(\lambda) =
T(\lambda)\ \tilde K(\lambda)\ T^{t}(-\lambda),
\label{treps}
\ee
$K,\ \tilde K$ are $c$-number
solutions of the reflection algebra and twisted Yangian respectively

Connection of this construction with the local boundary procedure described in section 2
is established by the following remark:
Consider equation (\ref{treps}), then take the linear limit, that is
\be
T(\lambda) \to 1 + \eta \int_{-L}^0 dx\  L(\lambda, x),\ ~~~~T^{\sigma}(\lambda) \to 1 +\eta
\int_{-L}^0 dx\ L^{\sigma}(\lambda, x), ~~~~~K(\lambda) \to {\mathrm k}
\ee
where $\sigma$ defined in (\ref{anti}) and (\ref{sigma2}) for $L$ and $T$ respectively. Then via (\ref{treps})
\be {\cal T}(\lambda) \to {\mathrm k} + \eta \int_{-L}^0 dx\ {\mathbb T}(\lambda, x) \ee
and (\ref{gen}) is the linear limit of (\ref{refc1}), (\ref{refc2}). Notice that we introduce the parameter $\eta$
to show that we only keep first order linear terms (see (\ref{mono2})).

Expansion of the representations in powers of $\lambda^{-1}$ will provide the boundary non-local charges of the PCM.
First the expansion of
$T(\lambda)= \sum_i{T^{(k)} \over \lambda^{k+1}}$ where the first terms are given as
\be
T^{(0)} =  \int_{-L}^0 dx\ j_{0}(x),  ~~~~T^{(1)} =  \int_{-L}^0 dx\ j_{0}(x) + \int_{0 \geq x > y\geq -L}dx dy\ j_{0}(x) j_{0}(y).
\ee
The asymptotic expansions of the $c$-number matrices are also given by
\be
K(|\lambda| \to \infty) \sim k
+ {1\over \lambda }f +{\cal O}({1\over \lambda^2}), ~~~~~
\tilde K(|\lambda| \to \infty) \sim \tilde  k +
{1\over \lambda }\tilde f +{\cal O}({1\over \lambda^2}).
\ee
\\
Then it is quite straightforward to see that the first terms of the expansions
\be
{\cal T}(\lambda) =\sum_k {{\cal T}^{(k)} \over \lambda^{k+1}} ,\ ~~~~
\tilde {\cal T}(\lambda) =\sum_k{\tilde {\cal T}^{(k)} \over \lambda^{k+1}}
\ee
which are the so-called boundary non-local charges, are given by:
\be
{\cal T}^{(0)}(0, -L) &=& \int_{-L}^0
dx \Big ( j_0(x) k +  k j_0(x) \Big ) \non\\  {\cal T}^{(1)}(0, -L) &= & \int_{-L}^0 dx \Big ( j_1(x)
 k - k j_1(x) \Big ) + \int_{-L}^0 dx_1 dx_2\ j_0(x_1) k j_0(x_2) +  \int_{-L}^0
dx \Big ( j_0(x) f +  f j_0(x) \Big )\non\\
&+&  \int_{0 \geq x_1 > x_2 \geq -L}
dx_1 dx_2\ j_0(x_1) j_0(x_2) k  + \int_{-L\leq x_1 < x_2 \leq 0} dx_1 dx_2\ k j_0(x_1) j_0(x_2). \non\\ \label{nl1}
\ee
See also \cite{dema} for the boundary PCM based on the reflection algebra.

Similarly for the twisted Yangian:
\be
\tilde {\cal T}^{(0)}(0, -L) &=& \int_{-L}^0 dx \Big ( j_0(x)\tilde k - \tilde k j^t_0(x) \Big ) \non\\
\tilde {\cal T}^{(1)}(0, -L) &= & \int_{-L}^0 dx \Big ( j_1(x)\tilde  k + \tilde  k j^t_1(x) \Big )-
\int_{-L}^0 dx_1 dx_2\ j_0(x_1)\tilde  kj^t_0(x_2) +\int_{-L}^0 dx \Big ( j_0(x)\tilde f - \tilde f j^t_0(x) \Big ) \non\\
&+&  \int_{0 \geq x_1 > x_2 \geq -L} dx_1 dx_2\ j_0(x_1) j_0(x_2)
\tilde  k  + \int_{-L\leq x_1 < x_2 \leq 0} dx_1 dx_2\ \tilde  k j^t_0(x_1) j^t_0(x_2).
\non\\ \label{nl2}
\ee
From these expressions it is clear that the continuum non-local charges (\ref{nl1}), (\ref{nl2}) are, as one would expect,
in exact correspondence with their discrete counterparts \cite{doikouy}.
The whole procedure is therefore at least self-consistent and nicely illustrates the use of both
non-local and boundary structures to yield physically relevant quantities.

It is clear in addition that the construction which we propose
can certainly be made systematic and universal, that is independent of the choice of model ($r,s$ matrix and Lax matrix)
\\
\\
{\bf Acknowledgments:}
We wish to thank Jean Michel Maillet for enlightening comments and suggestions.

\end{document}